\documentclass[12pt]{article}
\usepackage{a4,epsfig,amssymb}

  \newcommand{\ccaption}[2]{
    \begin{center}
    \parbox{0.85\textwidth}{
      \caption[#1]{\small{#2}}
      }
    \end{center}
    }

\newcommand{\beq}{\begin{equation}}
\newcommand{\eeq}{\end{equation}}
\newcommand{\beqa}{\begin{eqnarray}}
\newcommand{\eeqa}{\end{eqnarray}}

\newcommand{\Eqn}[1]{Eq.~(\ref{#1})}

\begin{document}

\pagestyle{empty}
\begin{flushright}
  IPPP/08/74 \\
  DCPT/08/148 \\
\end{flushright}
\vspace*{1cm}
\begin{center}
  {\sc \large 
The charm quark mass from non-relativistic sum rules } \\
   \vspace*{2cm}
{\bf  Adrian~Signer}\\
\vspace{0.6cm}
{\it Institute for Particle Physics Phenomenology \\
Durham, DH1 3LE, England \\}
  \vspace*{2.8cm} {\bf Abstract} \\ 
\vspace{1\baselineskip}
\parbox{0.9\textwidth}{ We present an analysis to determine the charm
  quark mass from non-relativistic sum rules, using a combined
  approach taking into account fixed-order and effective-theory
  calculations. Non-perturbative corrections as well as higher-order
  perturbative corrections are under control. For the PS mass we find
  $m_{{\rm PS}}(0.7~{\rm GeV}) = 1.50\pm 0.04$~GeV which translates
  into a $\overline{\rm MS}$ mass of $\overline{m} = 1.25\pm
  0.04$~GeV.}
\end{center}
\vspace*{5mm}
\noindent

\newpage

\setcounter{page}{1}
\pagestyle{plain}


\section{Introduction  \label{sec:I}}

In the sum rule approach~\cite{Novikov:1976tn} to determine the mass
$m$ of heavy quarks, $Q$, the sensitivity of the cross section
$\sigma(e^+e^-\to Q\bar{Q})$ near threshold $\sqrt{s}\simeq 2 m$ is
exploited by comparison of the experimental value of the $n$-th moment
$M_n$ to the theoretical prediction. The moments are defined as
\begin{equation}
M_n \equiv \int_0^\infty \frac{ds}{s^{n+1}}\, R_{Q\bar{Q}}(s)
= \frac{12 \pi^2 e_Q^2}{n!} 
\left(\frac{d}{d q^2}\right)^n \Pi(q^2)\big|_{q^2=0}
\label{MnDef}
\end{equation}
where $\Pi(q^2)$ is the vacuum polarization, $e_Q$ the electric charge
of the heavy quark and $R_{Q\bar{Q}}(s)\equiv \sigma(e^+e^-\to
Q\bar{Q})/\sigma(e^+e^-\to \mu^+\mu^-)$ the normalized cross section.
Traditionally, there have been two complementary theoretical
approaches to determine $M_n$. If $n$ is chosen to be small,
$n\lesssim 4$ the moments are evaluated using a fixed-order
approach. Sometimes this approach is referred to as relativistic or
low-$n$ sum rules. Alternatively, if $n$ is large, fixed-order
perturbation theory breaks down due to the presence of terms
$(\alpha_s \sqrt{n})^l$. In order to get a reliable theoretical
prediction these terms have to be resummed, counting $v\sim
1/\sqrt{n}\sim\alpha_s$, where $v$ is the (small) velocity of the
heavy quarks. This is usually done in an effective-theory approach
(for a review see Ref.~\cite{Brambilla:2004jw}), treating the heavy
quarks in the non-relativistic approximation. Therefore, this approach
is referred to as non-relativistic sum rules.

Relativistic sum rules have been used to determine the bottom and
charm quark masses~\cite{Kuhn:2001dm, Hoang:2004xm, Boughezal:2006px,
  Kuhn:2007vp}. The first two moments are known at four
loop~\cite{Boughezal:2006px, Chetyrkin:2006xg, Kniehl:2006bf,
  Maier:2008he}, i.e. ${\cal O}(\alpha_s^3)$, higher moments are
currently known at ${\cal O}(\alpha_s^2)$~\cite{Chetyrkin:1997mb,
  Boughezal:2006uu}. The extracted mass and its error depend crucially
on how the experimentally poorly known continuum cross section is
treated and how the theoretical error is estimated. Recently the charm
quark mass has been determined using this approach but replacing
experimental data for $\sigma(e^+e^-\to c\bar{c})$ by input from
lattice QCD~\cite{Allison:2008xk}, which via tuning uses different
experimental input such as the $\eta_c$ mass.

Applications of the non-relativistic sum rules have been restricted to
the determination of the bottom quark mass~\cite{BmassNR,
  Pineda:2006gx} so far. The moments are known to next-to-next-to
leading order (NNLO), in the counting of the effective theory, and in
the case of Ref.~\cite{Pineda:2006gx} also include the resummation of
logarithms~\cite{resumLog} of the form $(\alpha_s \ln\sqrt{n})^l$ at
next-to-leading (NLL) and partially at next-to-next-to-leading
logarithmic accuracy (NNLL). A complete NNNLO calculation is still
missing but partial results are available~\cite{ETnnnlo,
  Beneke:2008cr}. This method of extracting $m$ is virtually
insensitive to the continuum cross section but suffers from large
higher-order corrections.

The main reason why non-relativistic sum rules have not been used in
the case of charm quarks is that the application of perturbation
theory in this context is thought to be questionable. First, the
typical non-relativistic momentum and energy scales, $2 m/\sqrt{n}$
and $m/n$, are very small for large $n$. To some extent this is
related to the large higher-order corrections mentioned above and in
fact is already a problem for the case of bottom quarks. Second, the
non-perturbative contributions from vacuum condensates increase with
increasing $n$ and decreasing $m$ and potentially make a precise
determination of $M_n$ impossible.

For what values of $n$ can sum rules be used in the charm case and
when is perturbation theory not applicable any longer?  In order to
answer this question we will consider $M_n$ for all $n\le 16$ in the
charm case. We will use an approach that combines techniques for
low-$n$ and large-$n$ sum rules. In the case of the bottom quark it
has been shown~\cite{Signer:2007dw} that even though these techniques
are completely different, the results are remarkably consistent.
Encouraged by this we perform an all $n$ analysis in the case of the
charm quark. We will see that using a combined approach helps to keep
the size of higher-order corrections under control. Also, the
non-perturbative corrections due to the gluon condensate turn out to
be much smaller than expected, even for large $n$, if a threshold mass
definition~\cite{massdef, Beneke:1998rk, Pineda:2001zq} is used. This
will lead us to conclude that, contrary to common belief, the charm
quark mass can be extracted from non-relativistic sum rules in a
reliable way.

In Section~\ref{sec:D} we will describe how to obtain the experimental
moments, the theoretical moments and the non-perturbative
contributions in turn. We then combine these results in
Section~\ref{sec:R} and determine the charm quark mass.

\section{Determination of the moments \label{sec:D}}

\subsection{The experimental moment \label{sec:Dexp}}

We start with the determination of the experimental moments. This is
conveniently split into three regions: the resonance region including
the bound states $J/\Psi$ and $\Psi(2S)$ below threshold, the
threshold region $2 M_{D_0} = 3.73~{\rm GeV} \leq \sqrt{s} \leq
4.8~{\rm GeV}$ and the continuum region $\sqrt{s} > 4.8~{\rm GeV}$.

The mass and leptonic width of $J/\Psi$ and $\Psi(2S)$ are known to a
high accuracy which leads to a very precise determination of the
resonance contribution. We use $M_{J/\Psi} = 3096.916(11)$~MeV,
$M_{\Psi(2S)} =3686.09(4)$~MeV, $\Gamma_{J/\Psi} = 5.55(14)$~keV and
$\Gamma_{\Psi(2S)} = 2.38(4)$~keV~\cite{Amsler:2008zz}. The resonance
contribution is then given by
\begin{equation}
\label{Mresonance}
M_n^{\rm (res)} = \frac{9\pi}{\alpha^2} \sum_i \frac{\Gamma_i}{M_i^{2n+1}}
\end{equation}
where $i\in\{J/\Psi,\Psi(2S)\}$ and $\alpha = \alpha(M_i)= 1/134$.

The contribution from the threshold and continuum region are much more
difficult to determine. However, for increasing $n$ these
contributions become less and less important. In fact, for $n > 5$
the combined threshold and continuum contribution to the moments is
smaller than the error from the resonance contribution. Since our
analysis will be driven by large $n$ a rather crude determination of
these contributions with a large error will not affect the final
result.  This is one of the big advantages of this approach compared
to a fixed-order low-$n$ analysis. In particular, there is no need to
replace experimental data above threshold by theoretical input to
(artificially) decrease the experimental error.

For the continuum contribution we use data points above and below
threshold~\cite{Bai:2001ct} to obtain a very crude parameterization
$R_{c\bar{c}}(s) = 1.4\pm 0.5$ for $\sqrt{s} > 4.8$~GeV. In the
threshold region, we include $\Psi(3770)$, $\Psi(4040)$, $\Psi(4160)$
and $\Psi(4415)$~\cite{Amsler:2008zz} according to \Eqn{Mresonance} in
addition to an underlying contribution parameterized by
$R_{c\bar{c}}(s) = -4.88 + 1.31 \sqrt{s}$. This corresponds to a
linear in $\sqrt{s}$ extrapolation between
$R_{c\bar{c}}(\sqrt{s}=3.73~{\rm GeV}) = 0$ and
$R_{c\bar{c}}(\sqrt{s}=4.8~{\rm GeV}) = 1.4$. This is of course a very
crude estimate. We take this into account by taking as the error the
full size of the underlying contribution. With this procedure we
obtain the results as given in Table~\ref{tab:expmom} for the
experimental moments. The total error has been obtained by adding the
separate errors in quadrature.

\begin{table}[h]
\begin{center}
\bigskip
\begin{tabular}{|l|c|c|c||c|}
\hline
$n$ & $10^{n-1}M_n^{\rm res}$ & $10^{n-1}M_n^{\rm thr}$ &
$10^{n-1}M_n^{\rm cont}$ & $10^{n-1}M_n^{\rm exp}$ \\
\hline
1&0.1190(28)  &0.0361(176)  &0.0608(217)  & 0.2158(281)  \\
2&0.1167(28)  &0.0202(92)  &0.0132(47)  &0.1501(107)  \\
3&0.1162(28)  &0.0115(49)  &0.0038(14)  &0.1315(58)   \\
4&0.1171(29)  &0.0067(27)  &0.0012(4)  &0.1250(39)   \\
5&0.1192(30)  &0.0039(15)  &0.0004(1)  &0.1235(33)   \\
6&0.1221(30)  &0.0023(8)  &0.0002(1)  &0.1246(32)   \\
7&0.1257(31)  &0.0014(5)  &0.0001(1)  &0.1272(32)   \\
8&0.1299(33)  &0.0009(3)  &0  &0.1308(33)   \\
9&0.1346(34)  &0.0005(2)  &0  &0.1352(34)   \\
10&0.1397(35)  &0.0003(1)  &0  &0.1400(35)   \\
11&0.1452(37)  &0.0002(1)  &0  &0.1454(37)   \\
12&0.1510(38)  &0.0001(0)  &0  &0.1512(38)   \\
13&0.1572(40)  &0.0001(0)  &0  &0.1573(40)   \\
14&0.1637(41)  &0.0001(0)  &0  &0.1638(41)   \\
15&0.1706(43)  &0  &0  &0.1706(43)   \\
16&0.1778(45)  &0  &0  &0.1778(45)   \\
\hline
\end{tabular}
\end{center}
\ccaption{}{Values of experimental moments (in [GeV]$^{-2n}$) and
  their errors. Entries smaller than $5\times 10^{-5}$ are given as 0.
\label{tab:expmom} }
\end{table}

We stress that it is of course possible to get more precise results
for small $n$, but in our approach this is not required. In fact, in
what follows we will consider only $n>2$ and for our final result only
moments with $n\gtrsim 8$ are relevant.

\subsection{The theoretical moment \label{sec:Dth}}

The evaluation of the theoretical moments follows the discussion given
in Ref.~\cite{Signer:2007dw}. We consider three ways to evaluate the
moments: fixed-order (FO) moments, moments evaluated using an
effective-theory approach (ET) writing
\begin{equation}
M_n = \int_{-\infty}^\infty
 \frac{2\, dE}{(2 m)^{2n+1}}\, e^{-\frac{n E}{m}} 
\left(1 - \frac{E}{2 m} + \frac{n E^2}{(2 m)^2}  
  +\ldots \right) R_{c\bar{c}}(E)
\label{momET}
\end{equation}
with $E=\sqrt{s}-2m$ and, finally, moments using a combined
approach. The latter are obtained by adding the FO and ET moments and
subtracting the doubly counted terms~\cite{Signer:2007dw}. These
moments should provide us with a reliable theoretical prescription for
all $n$ as long a non-perturbative corrections are under control.

Since we are dealing (at least partially) with large $n$ where the
moments are dominated by the lowest lying resonances, we have to use a
mass definition adapted to the description of such resonances, i.e. a
threshold mass~\cite{massdef, Beneke:1998rk, Pineda:2001zq}. We use
the PS mass~\cite{Beneke:1998rk} with the associated factorization
scale $\mu_F=0.7$~GeV. For the strong coupling we set $\alpha_s(M_Z) =
0.118$ and use three-loop evolution.

The main issue for a reliable extraction of the charm mass will be a
realistic estimate of the theoretical error due to missing
higher-order corrections. This is a notorious problem and there is no
generally applicable procedure.  We will therefore use a combination of
different methods and criteria, as described below.

\begin{figure}[h]
   \epsfxsize=13cm
   \centerline{\epsffile{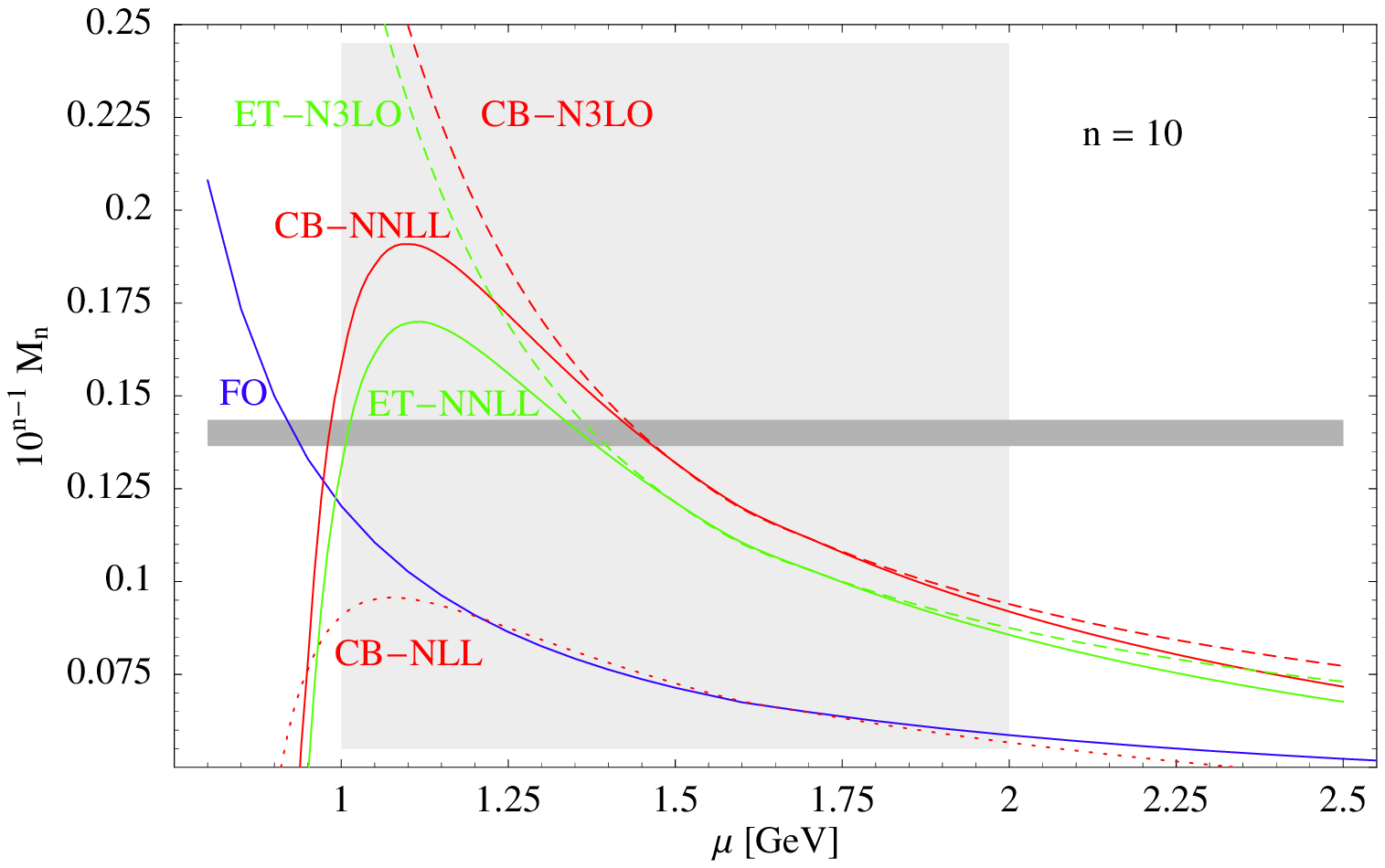} }
   \vspace*{0.2cm} \ccaption{}{Scale dependence of various theoretical
     predictions for $M_{10}$ evaluated with $m_{\rm PS}(0.7~{\rm
       GeV}) = 1.50$~GeV. $M_{10}^{\rm exp}$ is shown as thin grey
     band and the range of scale variation used for $\delta m^{\rm
       th}$ is indicated by the light-grey region.
 \label{fig:CharmMom10}}
\end{figure}

Let us use the 10th moment as an example to illustrate our
determination of the theoretical error. In Figure~\ref{fig:CharmMom10}
we display $M_{10}$ as a function of the scale $\mu$ for $m_{\rm
  PS}(\mu_F) = 1.50$~GeV. We consider a range of different theoretical
predictions.
\begin{description}
\item{FO:} this is a fixed-order calculation including all
  terms up to ${\cal O}(\alpha_s^3)$, keeping in mind that the
  constant term of  ${\cal O}(\alpha_s^3)$ is actually not yet known
  for $n>2$. We have fixed this constant to be the one of the first
  moment. This has only a very weak influence on the result.
\item{ET-NNLL:} this is a renormalization-group improved
  effective-theory calculation complete at NNLO. The resummation of
  logarithms is complete at NLL and partially done at NNLL. For
  details we refer to Ref.~\cite{Pineda:2006gx}. 
\item{ET-N3LO:} this is an effective-theory calculation where the
  logarithms have been re-expanded and kept up to NNNLO. This result
  contains all terms at NNLO and the logarithmically enhanced NNNLO
  terms. It is used to gauge the impact of corrections beyond NNLO and
  the importance of resumming the logarithms.
\item{CB-NNLL:} this is a combined FO-ET
  calculation~\cite{Signer:2007dw}. The ET input is as for ET-NNLL. In
  addition all terms of ${\cal O}(\alpha_s^3)$ have been
  included. This is the ``best'' available theoretical prediction.
\item{CB-N3LO:}  this is also a combined FO-ET calculation,
  but the ET input has been taken as in ET-N3LO.
\item{CB-NLL:}  this is also a combined FO-ET calculation,
  but the ET input is modified to be only at NLO/NLL. The subtraction
  to avoid double counting when combining with the FO result has to be
  adapted accordingly. This result is used to consider the convergence
  of the perturbative series.
\end{description}

The following points related to Figure~\ref{fig:CharmMom10} will be
of importance and in fact are valid for all moments that are relevant
to us, i.e. with $n\gtrsim 5$:
\begin{itemize}
\item The ET and CB results are very close, indicating that the
  relativistic corrections to the ET result are small. This is not
  surprising for large $n$. What is surprising to some extent is that
  the relativistic corrections turn out to be small also for small $n$.
\item The ET-NNLL and CB-NNLL results have a peak slightly above
  $\mu\simeq 1$~GeV. For scales below the peak, the results become
  very soon unreliable indicating a breakdown of perturbation
  theory. The peak is close to $\mu=2 m/\sqrt{n}$, the typical
  momentum scale in the non-relativistic region.
\item The ET-N3LO and CB-N3LO results are very similar to the ET-NNLL
  and CB-NNLL  results except for small $\mu$. It is in this region
  only, where the resummation of logarithms actually becomes
  important and, therefore, the ET-N3LO and CB-N3LO results cannot be
  used any longer.
\item The FO prediction does remarkably well even for large moments,
  where it is supposed to be inapplicable. This hinges on the fact
  that a threshold mass has been used in the FO approach.
\end{itemize}

Taking into account these observations we proceed as follows to
determine the mass and its theoretical error due to missing
higher-order corrections. We start by taking our best prediction,
CB-NNLL, and determine a band of $m$ values by varying $1~{\rm GeV}
\le \mu \le 2~{\rm GeV}$. The standard prescription would be to vary
the scale by a factor two around the typical value $\mu = 2
m/\sqrt{n}$. This would result in scales below 1~GeV though, which
according to Figure~\ref{fig:CharmMom10} are not acceptable. However,
if the upper limit of the standard variation is larger than 2~GeV
(i.e. for $n\le 9$) we use the larger value instead. Note that in any
case the peak of the CB-NNLL result is included in the band of scale
variation. We now extract the mass as the central value of the band
with symmetric errors. The results are shown in the first column of
Table~\ref{tab:theoerr}.  As anticipated small moments have a large
error in our approach and will not play a significant role.

\begin{table}[h]
\begin{center}
\bigskip
\begin{tabular}{|r|c|c||c||c|c|}
\hline 
  $n$ &CB-NNLL & CB-N3LO &$m(\delta m^{\rm th})$&
 CB-NLL & FO ${\cal O}(\alpha_s^3)$ \\
\hline
3 & 1436(156) & 1583(182) &1509(229) & 1451 &1434  \\
4 & 1464(102) & 1553(136) &1508(147) & 1439 &1434  \\
5 & 1478(72) & 1539(107)  &1509(103) & 1438 &1438  \\
6 & 1483(57) & 1531(88)   &1507(81)  & 1444 &1442  \\
7 & 1488(45) & 1526(74)   &1507(64)  & 1448 &1447  \\
8 & 1493(36) & 1524(63)   &1508(52)  & 1452 &1452  \\ 
9 & 1494(30) & 1521(55)   &1508(44)  & 1456 &1457  \\
10& 1494(28) & 1518(50)   &1506(40)  & 1463 &1460  \\
11& 1494(26) & 1516(46)   &1505(37)  & 1466 &1463  \\
12& 1494(25) & 1514(43)   &1504(35)  & 1470 &1467  \\
13& 1494(23) & 1513(40)   &1503(32)  & 1473 &1470  \\
14& 1495(22) & 1511(38)   &1503(31)  & 1479 &1473  \\
15& 1495(21) & 1511(36)   &1503(29)  & 1481 &1476  \\
16& 1496(20) & 1510(33)   &1503(27)  & 1484 &1479  \\
\hline    
\end{tabular}
\end{center}
\ccaption{}{Extracted mass and theoretical error in MeV, using various
  approaches. The central column shows the combined result with error.
\label{tab:theoerr} }
\end{table}

Given the remarkable consistency and the small errors of these results
it would be tempting to simply take them at face value. However, the
scale dependence alone is a dubious way to determine the theoretical
error. In order to get a more reliable estimate we extend the
analysis. We repeat the same exercise for the CB-N3LO case. The
results are given in the second column of Table~\ref{tab:theoerr}. As
can be seen from Figure~\ref{fig:CharmMom10} the CB-N3LO calculation
leads to larger moments (for small $\mu$) and therefore somewhat
larger values for $m$. Also, the error is dominated by small scales
$\mu\simeq 1$ where the CB-N3LO starts to become unreliable due to the
importance of resumming logarithms.  Therefore, taking the upper end
of the $m$ band of the CB-N3LO results in an overestimate of the
theoretical error. We thus combine the CB-NNLL and CB-N3LO results by
subtracting the CB-NNLL error from the (smaller) CB-NNLL result and
adding it to the (larger) CB-N3LO result. From this range we determine
$m$ (as the central value) and symmetric errors. In this way we take
into account the CB-N3LO tendency to give larger values of $m$ while
discarding the unreliable small scale region of the CB-N3LO
results. The results are given in the third column of
Table~\ref{tab:theoerr}.

Finally we perform two further cross checks on our error. We determine
the mass using the CB-NLL calculation and using the same scale
variation as above. The central value of the results are shown in
column~4 of Table~\ref{tab:theoerr}. These values all lie within the
error band which gives further confidence in our results. The same is
even true for the mass values determined by a FO approach, listed in
the last column of Table~\ref{tab:theoerr}. This could be taken as an
indication that the error has been overestimated. However, 
the corrections in the ET approach are very large and the NLL
results lie within the NNLL error band only because they have been
improved using the FO results in a combined analysis. Thus we prefer to
keep the larger error, anticipating relatively large NNNLO corrections
in the effective theory.

\begin{figure}[h]
   \epsfxsize=13cm
   \centerline{\epsffile{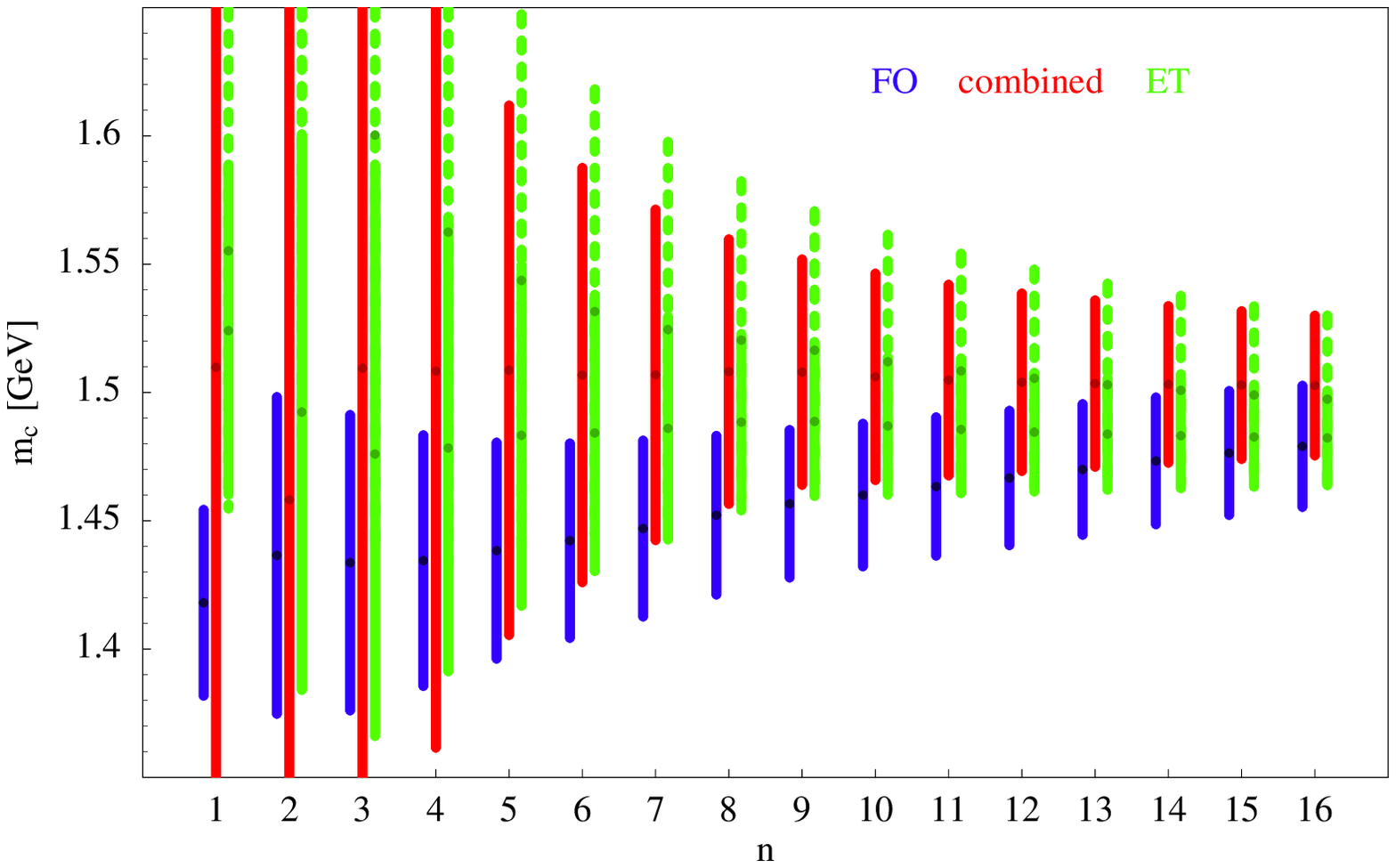} }
   \vspace*{0.2cm} \ccaption{}{Extracted values with theoretical
     errors for $m_{\rm PS}(0.7~{\rm GeV})$ for all moments $n\le 16$,
     using a FO (left, dark blue bands), an ET (right, light
     green bands) and a combined approach (middle, red bands).
 \label{fig:mcA}}
\end{figure}

To visualize the consistency of our approach, in Figure~\ref{fig:mcA}
we plot the extracted mass with its error for all $n$. The left (dark
blue) bands show $m_{\rm PS}$ as extracted using a FO approach with
the central value indicated by a dot. The right (light green) bands
show the corresponding values using a ET-NNLL and a ET-N3LO
approach. The latter leads to slightly larger values and errors for
$m_{\rm PS}$ and is depicted by the dashed band. The two dots in this
band indicate the two central values for ET-NNLL (lower) and ET-N3LO
(higher) respectively. Finally, the middle (red) bands show our
combined result, as given in in the third column of
Table~\ref{tab:theoerr}, with the central value again indicated by a
dot.

As expected, the central value of the combined result is close to the
ET result for large $n$. For small $n$, the combined result is also
consistent with the FO result, at the price of having a huge error.
The combined result makes use of all available information and gives
the most reliable prediction for large $n$.

\subsection{The non-perturbative contribution \label{sec:Dnp}}

One of the main reasons why non-relativistic sum rules so far have not
been used to extract the charm quark mass is the common belief that
non-perturbative corrections are not under control. As we will see
this is actually not the case.

A first hint that the situation is in much better control than
anticipated is the fact that the (central value of the) mass
extracted, as indicated by the points in Figure~\ref{fig:mcA} is
remarkably consistent for all values of $n\ge 5$. If there were large
non-perturbative effects they would with all likelihood affect results
with increasing $n$ more dramatically. 

In order to get a more quantitative picture, we will consider the
effects of the gluon condensate~\cite{Shifman:1978bx} which gives us a
handle for the leading non-perturbative correction. The corresponding
contribution to the sum rule has been computed to two
loops~\cite{Broadhurst:1994qj} and reads
\begin{equation}
\label{GlueCond}
\delta M_n^{({\rm np, X})} = \frac{12 \pi^2 e_Q^2}{(4 m^2)^{n+2}}\,
\langle \frac{\alpha_s}{\pi} G^2 \rangle \, 
a_n\, \left(1+\frac{\alpha_s}{\pi} 
\left[b_n - (2n+4)\, \delta b_X\right]\right)
\end{equation}
where the one-loop coefficients are given by
\begin{equation}
a_n = -  \frac{(2n+2)}{15} \frac{\Gamma(4+n)}{\Gamma(4)}
 \frac{\Gamma(7/2)}{\Gamma(7/2 + n)}
\label{an}
\end{equation} 
and $b_n$ are the two-loop coefficients in the pole scheme, as listed
in Ref.~\cite{Broadhurst:1994qj}. The shifts $\delta b_X$ take into
account the change in the mass scheme. For the PS mass we have
\begin{equation}
b_n^{\rm PS} \equiv b_n - (2n+4)\,\delta b_{\rm PS} = 
b_n - (2n+4)\, C_F \, \frac{\mu_F}{m}
\label{delbPS}
\end{equation}
where $C_F=4/3$ is a colour factor.

There are two issues we have to consider: how large are the
contributions due to the gluon condensate and how reliable is the
prediction in \Eqn{GlueCond}? The answer to both questions crucially
depends on the mass scheme used. Regarding the former, the key
features of \Eqn{GlueCond} are that the coefficients grow like
$a_n\sim n^{3/2}$ and that $\delta M_n^{({\rm np})}$ increases rapidly
for decreasing mass. To assess the situation, we calculated for the
case of the PS scheme the ratio of the non-perturbative contributions,
as given in \Eqn{GlueCond}, to the experimental moment, using $m_{\rm
  PS} = \mu =1.5$~GeV and $\langle (\alpha_s/\pi) G^2 \rangle =
0.005\, {\rm GeV}^4$~\cite{Ioffe:2005ym}. The results are shown in
Table~\ref{tab:npcontribution}.  As can be seen, the non-perturbative
contributions are well below 10\% for all $n\le 16$ indicating that
they are in fact not (yet) very important. Another satisfactory
feature of the PS scheme is that the series in \Eqn{GlueCond} is well
behaved. To show this we also list the relative importance of the
two-loop corrections. Clearly, they should be small compared to the
leading term, for \Eqn{GlueCond} to be applicable. For the first
moment, the higher order correction is 75\% of the leading term which
is uncomfortably large. However, for larger values of $n$, where the
contribution starts to become somewhat more relevant, the corrections
are smaller.

\begin{table}[h]
\begin{center}
\bigskip
\begin{tabular}{|l|c|c|c|c|c|c|}
\hline 
 & $n=1$  & $n=4$  & $n=7$  & $n=10$  & $n=13$  & $n=16$ \\
\hline
$\delta M_n^{({\rm np, PS})}/M_n^{\rm exp}\ [10^{-2}]$ 
& 0.1  & 0.7 & 1.6 & 2.9 & 4.3 & 5.9\\
$\alpha_s\, b_n^{\rm PS} / \pi$
& 0.75 & 0.72 & 0.61 & 0.46 & 0.28 & 0.09 \\
\hline
\end{tabular}
\end{center}
\ccaption{}{Importance of gluon condensate contribution to moments
  (first row) and relative importance of two-loop corrections to gluon
  condensate contribution (second row) in the PS scheme.
\label{tab:npcontribution} }
\end{table}

From the results in Table~\ref{tab:npcontribution} we conclude that
the gluon condensate contributions are under control for all values of
$n$ considered here. This seems to be in contradiction with what is
commonly stated. However, we stress that the picture is completely
different if either the pole mass or the $\overline{\rm MS}$ mass is
used. It is well known that there is a close interplay between vacuum
condensates and mass definitions~\cite{Ioffe:2002be, Ioffe:2005ym}. In
the case of the pole mass, the corrections are also relatively small
(mainly because $m_{\rm OS} > m_{\rm PS}$) but the series in
$\alpha_s$ in \Eqn{GlueCond} is completely unreliable because the
two-loop corrections exceed the one-loop corrections for all $n$. In
the case of the $\overline{\rm MS}$ mass, the contributions are huge
for large $n$ (mainly because $m_{\overline{\rm MS}} < m_{\rm PS}$)
and the corrections are also very large, unless extremely small scales
$\mu \lesssim 1$~GeV are used. For other threshold masses, such as
e.g. the RS mass~\cite{Pineda:2001zq} we checked that the main
conclusions are the same as for the PS mass. 

Of course one might wonder about the contribution of further
suppressed condensates such as the dimension 6 operator $\langle
G^3\rangle$. However, as we will see, the contribution and induced
error due to the $\langle G^2\rangle$ operator is so small that we can
simply take this into account by increasing the error. Thus we
conclude that if a mass definition adapted for quark pairs near
threshold is used, the non-perturbative corrections are under control
even in the charm case. We remark that this is also in agreement with
a recent analysis~\cite{Allison:2008xk} where contributions from the
gluon condensate were found to be much smaller than expected.

\section{Results \label{sec:R}}

In this section we extract the PS charm quark mass and determine the
various errors. The dominant error will be the error $\delta m^{\rm
  th}$ due to missing higher-order corrections discussed in
Section~\ref{sec:Dth}. We will consider all $3\le n\le 16$ even though
from Figure~\ref{fig:mcA} it is clear that values $n\lesssim 5$ are
``useless'' in the sense that their error is too large. The results
are summarized in Table~\ref{tab:mc}.  

The first column shows the central value for the mass. These entries
differ slightly from the corresponding entries of
Table~\ref{tab:theoerr} because the effect of the gluon condensate, as
discussed in Section~\ref{sec:Dnp}, has been included. Apart from the
theoretical error, taken directly from Table~\ref{tab:theoerr}, we
include three further sources of errors. 

First we consider the experimental error, $\delta m^{\rm exp}$. We
simply vary the experimental moments in the range given in
Table~\ref{tab:expmom} and consider the effect on the extracted
mass. As expected, the error decreases rapidly for increasing $n$ and
becomes very soon negligible.

A more relevant source of error is the uncertainty in the strong
coupling. We vary $0.116 \le \alpha_s(M_Z) \le
0.120$~\cite{Amsler:2008zz}. The resulting error, $\delta m^{\alpha}$,
is listed again in Table~\ref{tab:mc}.

Finally we consider the contribution and induced error due to the
gluon condensate, $\delta m^{GG}$. As discussed in
Section~\ref{sec:Dnp}, this contribution is surprisingly small. To
determine the error we vary $\langle (\alpha_s/\pi) G^2 \rangle =
0.005\pm 0.004\, {\rm GeV}^4$~\cite{Ioffe:2005ym} and determine the
corresponding change in $m$. We then double this error to take into
account higher-order corrections to \Eqn{GlueCond}, higher dimensional
vacuum condensates and the fact that previous determinations of
$\langle (\alpha_s/\pi) G^2 \rangle$ resulted in somewhat larger
values. Even so, the error does virtually not affect the final result.

The total error, listed in the last column of Table~\ref{tab:mc} is
obtained by adding the various errors in quadrature. We also checked
that the higher-order QED contributions have a negligible effect. In
fact, they change the mass by a few MeV at most.

\begin{table}[h]
\begin{center}
\bigskip
\begin{tabular}{|l|c||c|c|c|c||c|}
\hline
$n$ & $m$ & $\delta m^{\rm th}$ & $\delta m^{\rm exp}$
& $\delta m^{\alpha}$ & $\delta m^{GG}$ & $\delta m$  \\
\hline
3 &1508  &229  &11  &41  &2  &233  \\
4 &1507  &147  &6  &34  &3  &151  \\
5 &1508  &103  &4  &29  &3  &107  \\
6 &1506  &81  &3  &27  &3  &85  \\
7 &1505  &64  &3  &24  &4  &69  \\
8 &1506  &52  &2  &22  &4  &57  \\
9 &1504  &44  &2  &20  &4  &49  \\
10&1503  &40  &2  &19  &5  &45  \\
11&1503  &37  &2  &18  &5  &41  \\
12&1501  &35  &2  &17  &5  &39  \\
13&1500  &33  &1  &16  &5  &36  \\
14&1500  &31  &1  &15  &6  &35  \\
15&1500  &29  &1 &14  &6  &33  \\
16&1500  &27  &1  &14  &6  &31  \\
\hline
\end{tabular}
\end{center}
\ccaption{}{Extracted charm quark mass with separate and total
  errors. All entries are in MeV and for the PS mass with $\mu_F =
  0.7$~GeV.
\label{tab:mc} }
\end{table}

The extracted mass is virtually independent of $n$. Thus, the only
issue regarding how to combine the results of Table~\ref{tab:mc} is
the determination of the final error. Given the remarkable consistency
between the various results we argue it is safe to take a single
moment result with a rather large $n$. Therefore we take as our final
result
\begin{equation}
\label{PSres}
m_{\rm PS}(0.7~{\rm GeV}) = 1.50 \pm 0.04\, {\rm GeV}
\end{equation}
Since the determination of the dominant error, $\delta m^{\rm
  th}$ is somewhat arbitrary, we think it is misleading to give more
significant figures in the error.

Converting this to the $\overline{\rm MS}$ mass we obtain
$\overline{m} \equiv m_{\overline{\rm MS}}(m_{\overline{\rm MS}}) =
1.25$~GeV. The error of 40~MeV in \Eqn{PSres} results in an error of
35~MeV for the $\overline{\rm MS}$ mass. However, there is also an
error in the conversion itself. As an indication of this error we take
the size of the fourth order term in the conversion and obtain an
additional error of 15~MeV. Finally, there is an error in the
conversion induced by the uncertainty in $\alpha_s$. Varying $0.116
\le \alpha_s(M_Z)\le 0.120$ in the conversion and taking into account
the correlation of this with the corresponding variation in the
determination of the PS mass, this results in an error of 15~MeV as
well. These errors are relatively large because the coupling is large
due to the small scale. Thus a reduction in the error in \Eqn{PSres}
would only partially impact on the error in the $\overline{\rm MS}$
mass. Combining in quadrature the three errors in the conversion we
obtain for the $\overline{\rm MS}$ mass
\begin{equation}
\label{MSres}
\overline{m} = 1.25 \pm 0.04 \, {\rm GeV}
\end{equation}
This value is in good agreement with the world average $\overline{m} =
1.27^{+0.07}_{-0.11}$~\cite{Amsler:2008zz} but has a larger error than
recent determinations using low-$n$ moments~\cite{Boughezal:2006px,
  Kuhn:2007vp, Maier:2008he}. However, we would argue that our
estimate of the theoretical error is more conservative and that this
determination of the charm quark mass is in many respects
complementary to the low-$n$ sum rules.

\section{Conclusions \label{sec:C}}

The main result of this work is that the non-relativistic sum rule can
be used to obtain a precise and reliable determination of the charm
quark mass. The non-perturbative corrections are under control even
for $n \simeq 8 - 16$ as long as a suitable threshold mass definition
is used. The situation with respect to the large corrections in the
effective theory is much improved if a combined analysis is performed,
including available fixed order results.

We are aware that these statements are to a certain extent in
contradiction with what would naively be expected. However, looking at
the situation more carefully, they are actually not that
surprising. It has been shown previously~\cite{Signer:2007dw} that in
the case of the bottom quark, the FO as well as the ET approach work
much better than expected. In the charm case large moments were also
found to give consistent results~\cite{Allison:2008xk}.  The value of
the quark mass does not seem to be the driving force for the large
corrections in the effective theory. In fact, the corrections are also
large in the top case. Thus, the reduction in mass from bottom to
charm does not completely alter the question regarding the
applicability of perturbation theory. Given that completely different
theoretical approaches give comparable results and that the size of
the corrections are reasonable in a combined approach, we argue that
the situation regarding non-relativistic sum rules in the charm case
is similar to the bottom case. In spite of large partial NNNLO
corrections to the sum rules in the bottom case~\cite{Beneke:2008cr}
we expect that the total NNNLO correction is within our error
estimate, implying similar cancellation between the various NNNLO
contributions as for the top case. With a careful, conservative error
estimate the quark mass can be determined reliably.

The by far largest contribution to the error in the present
determination of the charm quark mass comes from unknown higher-order
corrections. An estimate of this error is notoriously difficult and to
a large extent arbitrary. It is for this reason that we deliberately
refrained from pushing the error estimate to an extreme. In
particular, to make our error estimate as reliable as possible, we do
not take the considerably smaller errors of the CB-NNLL result, nor do
we take the smallest error in Table~\ref{tab:mc}.

It is clear that neither a FO nor a ET analysis alone can cover the
whole range of $n$ and only a combined analysis can make use of all
available information. In this sense the present analysis can be
considered as to a large extent complementary to low-$n$ sum rules,
since it is clearly dominated by a large-$n$ approach. This approach
uses a different theoretical input and the consistency of this result
with other determinations provides useful information.

\vspace*{0.5em}
\noindent
\subsubsection*{Acknowledgement}
It is a pleasure to thank Roman Zwicky for many useful discussions.
This work is supported in part by the European Community's Marie-Curie
Research Training Network under contract MRTN-CT-2006-035505 `Tools
and Precision Calculations for Physics Discoveries at Colliders'.


\end{document}